# The Primary Design of the Ridgetron

Li Jinhai[#], Li Chunguang

China Institute of Atomic Energy

**Abstract:** The ridgetron is used to accelerate the intense beam for the electron irradiation. According to the electromagnetic field simulation, the extend electrode should be cancelled to increase the shunt impedence. The method of parameter sweeping with constraint variables (PSCV) is used to scan the mechanical parameters of the cavity to optimize the shunt impedence. The thermal analysis also has been done for water cooling. The beam dynamic simulation has been done at last.

**Key words:** ridgetron, electron irradiation, parameter sweeping with constraint variables (PSCV), thermal analysis, beam dynamic

**PACS**: 29.20.-c,29.20.Ba

## 1 Introduction

The electron irradiation technology is a high efficiency, ecotypic, secure technology. It can be used in the sterilization of medical disposables, the preservation of foods, and so on. The Ridgetron is the new type of high power and energy accelerator used for the electron irradiation field[1-3], which can accelerate the continuous electron beams of 1-25 kW in the 0.5-10 MeV range in a compact space.

There are two kinds of accelerator for the high energy electron irradiation: the linear accelerator and the recirculating accelerator. The electron linear accelerators (linac) have been widely used in many irradiation facilities. However, they are operated in pulsed mode and at low energy efficiency. The recirculating electron accelerator is a kind of cw accelerator, which include Rhodotron[4-6], Ridgetron, Fantron[7-9], and so on.

Noriyosu has designed and manufactured a small ridged cylindrical cavity for a beam power of 6.5 kW and a final energy of 2.5 MeV. The design quality factor of the cavity is about 27000. In order to improve the RF porformance, the ridgetron is designed and optimized in this article.

## 2 Ridged cavity model

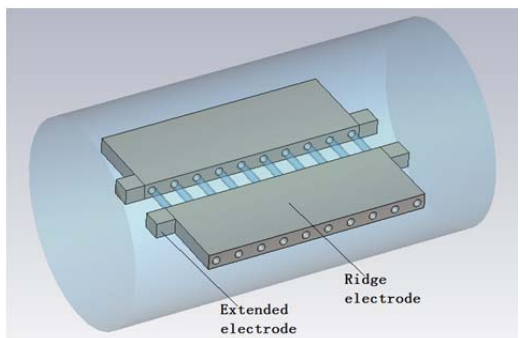 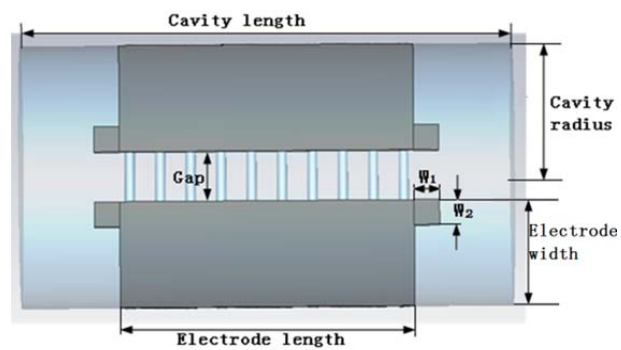

Fig.1 the three dimensional cavity    Fig.2 the parameters of the cavity

The ridged cavity is a cylindrical cavity equipped with two ridges as shown in Fig.1. In the ridged cavity, the electric field excited in the $TE_{110}$ mode is concentrated in the gap between two ridges and the magnetic field is zero in their gap. The electron is accelerated in the plane through the gap and inside the hollow ridges without parasitic deflection by the magnetic field. The parameters of the cavity are shown in Fig.2.

The principle of the ridgetron is shown in Fig.3. The electron beam is first produced by electron gun,

*Work supported by NSTS (2014BAA03B01) and CNNC (FA14000104)
#lijinhai@ciae.ac.cn

and then injected into the cavity. The beam is accelerated in the gap between the two ridge electrodes. When the beam output the cavity, it is deflected 180 degrees by magnets and injected into the cavity again. The same proceduce is recircled until the finial energy is obtained.

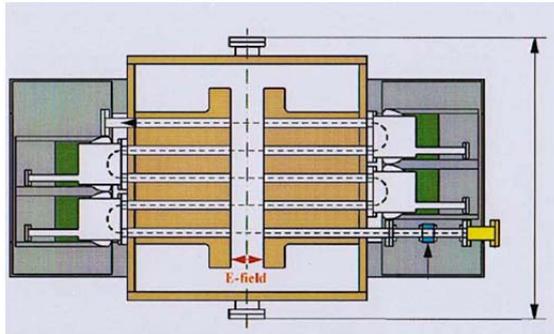

Fig. 3 the acceleating principle of the ridgetron

### 3 The influnce of the extended electrode

In order to uniform the electrical field in the gap, the extended electrode is used[1]. However the effect of the extended electrode on the other RF parameters should be taked into account. In table1, the cavity length (CL) is fixed at 2040mm, the cavity radius (CR) is fixed at 512mm, the Electrode Length (EL) is fixed at 1220mm, and the gap is fixed at 200mm. For Case2, W2 is 412mm, which means that the EL extend to 1420mm, at the same time the extended electrode is cancelled. For Case 1 with extended electrode, the quality value (Q) and the Shunt impedence is lower than that of Case 2 and Case 3.

Table 1 the comparision of the RF performance with and without extended electrode

| Case | W1(mm) | W2(mm) | Q | Shunt impedence(MΩ) |
| --- | --- | --- | --- | --- |
| 1 | 100 | 100 | 31000 | 4.56 |
| 2 | 100 | 412 | 32300 | 5.08 |
| 3 | 0 | - | 34000 | 5.63 |

For the first time acceleration at the first beam pipe, the electric field distribution along the beam line with and without the extended electrode is shown in Fig.4. Because the RF power dissipate on the cavity wall is same for the two cases, the more energy gain can be obtained for the case without extended electrode. At the same time, the energy gain without extended electrode is great than that with extended electrode at other beam pipe as shown in Fig.5. The uniformity of acceleration field along the redges with extended electrodes is better than that without extended electrodes[2,3], however, the effect of the uneven field on the beam dynamics can be neglected, because the acceleration at each time is independence.

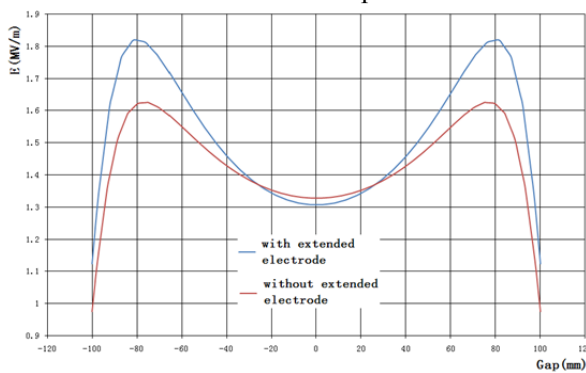
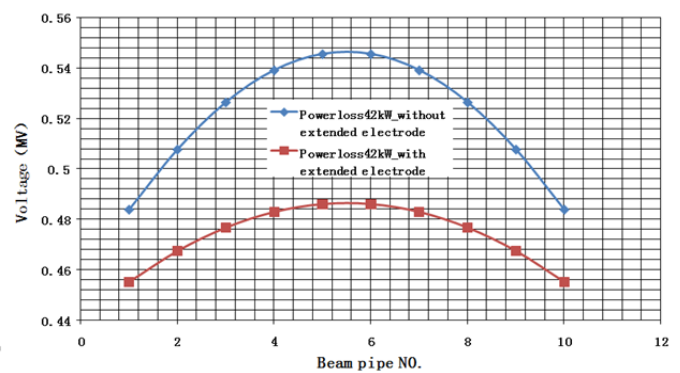

Fig. 4 the electric field distribution          Fig.5 the energy gain at different position

### 4 The cavity parameters scanning

During the parameters scanning, the method of "parameter sweeping with constraint variables"[10] (PSCV) is used for searching the optimum mechanical parameters at the fixed resonance frequency. The CR is used as a "constraint variable" to fix the resonant frequency to 100 MHz during the parameters scanning, because the resonant frequency is more sensitive to CR than it is to other geometrical parameters.

The optimise object is the shunt impedence, and the CL and Gap are scanded for the optimization as shown in Fig.6 and Fig.7. According to Fig.6 and Fig.7, the shunt impedence increase monotonously with CL and Gap. As we know, the CL and Gap can not increase unlimited and too large, so it is necessary to balance between the shunt impedence and the cavity mechanical parameters. And then, the parameters of the cavity is selected as: CL=2040mm, Gap=200mm, CR=512mm, EL=1220mm.

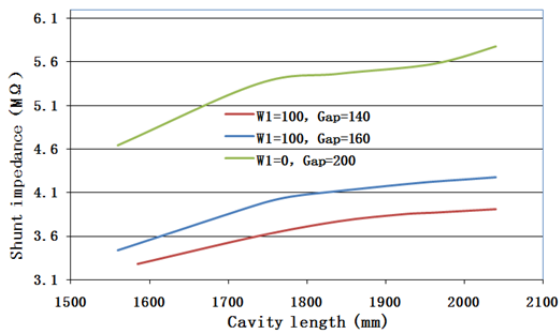 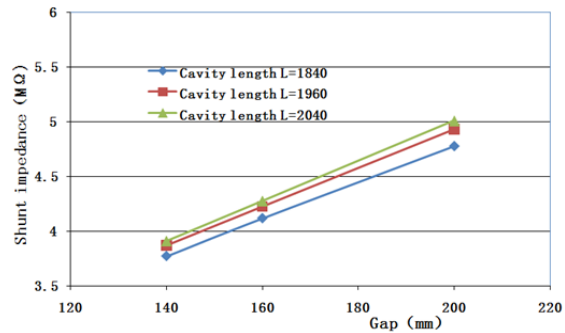

Fig.6 relation between CL and shunt impedence    Fig.7 relation between Cap and shunt impedence

## 5 the thermal analysis

During the electromagnetic field simulation, the surface current and power dissipate can be obtain as shown in Fig. 8 and Fig.9. It can be see that the end of the ridge electrode has more current density and power dissipate density, although more power is dissipated on the cylinder wall. So the more cooling at the end of the ridge electrode is design, and the maximum temperature can be controled under 80 Celsius degree when the temperture of the input cooling water is 25 Celsius degree as shown in Fig.10.

Table 2    the power dissipate at the cavity wall

| Total power | \multicolumn{2}{c}{50kW} ||
|---|---|---|
| | ratio | Power dissipate（w） |
| Cylinder wall | 57.90% | 28950 |
| | | 7237.5 |
| Ridge electrode 1 | 21.05% | 10525 |
| Ridge electrode 2 | 21.05% | 10525 |

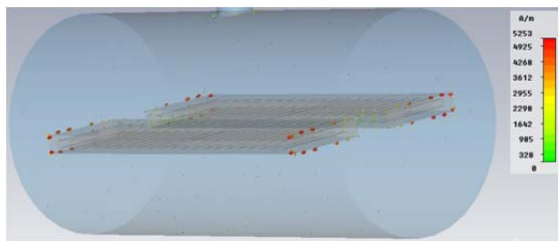 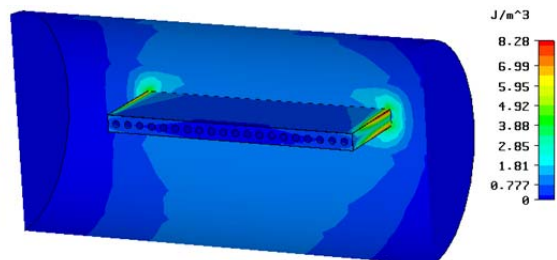

Fig.8 the surface current on the cavity wall        Fig. 9 the power dissipate on the cavity wall

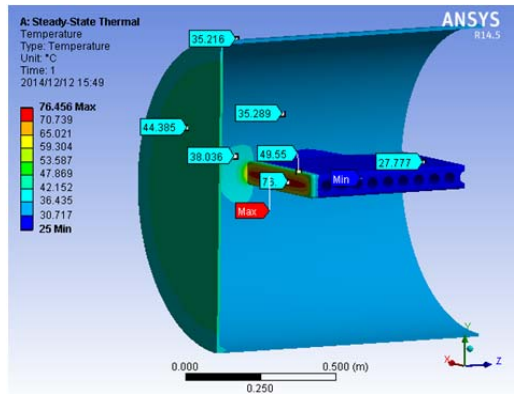

Fig.10 the thermal analysis with the cooling water

## 6 Beam dynamics

Because the energy gain is about 0.5MeV at each acceleration, 20 times accelerations in the cavity are needed, and then 19 bend magnets are needed. The 180 degree bend magnet[1] is divided three parts for dynamics simulation: one parts of main bending field and two parts of inverse bending field. The Trace3d code is used for the dynamics simulation. The pulsed current of each micro bunch is 60 mA, and the extract voltage of the electron gun is 20kV. Although the ridgetron accelerate the CW electron beam, the beam should bunched for decreasing beam loss during the RF field acceleration. If the bunch length is about 36 degree, the average current should be 6mA when the pulsed current is 60 mA. Because of the intense space charge force, the envelope expand to 35mm before the beam inject into the cavity with the energy of 20keV. The envelope can be control under 20mm during the later acceleration in Fig.11.

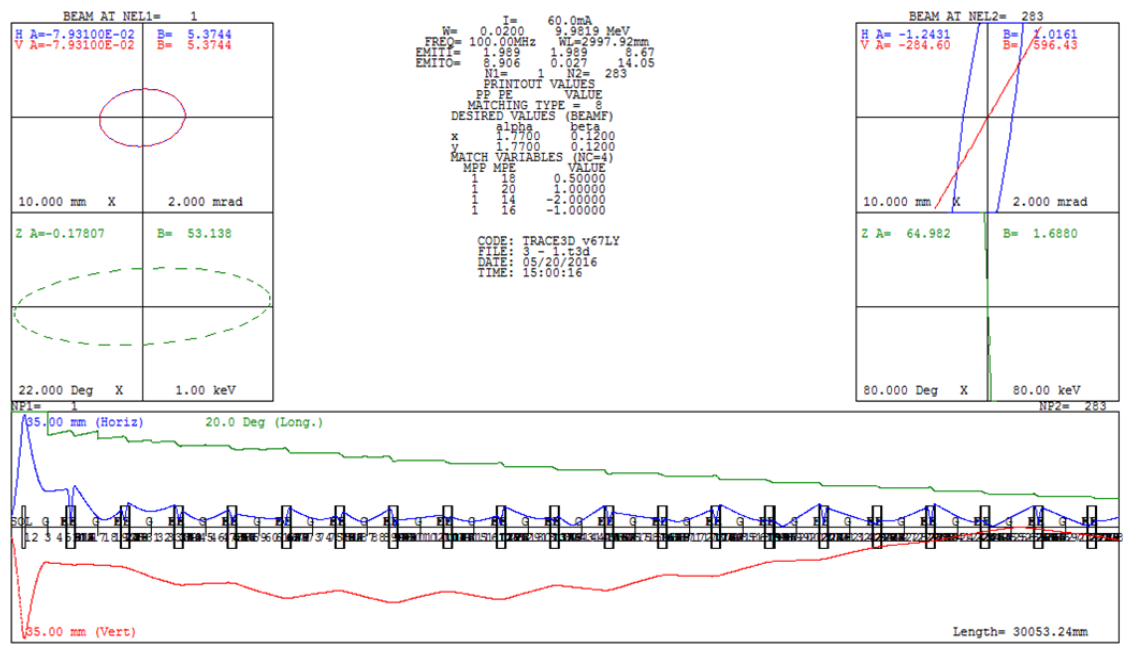

Fig. 11 the beam envelope of ridgetron

## 7 Conclusion

According to the electromagnetic field simulation, the extend electrod should be cancelled to improve the shunt impedence. Although the method of parameter sweeping with constraint variables is used, the maximum can not be found when CL and Gap are scanned, and then it is necessary to balance between the shunt impedence and the cavity mechanical parameters. Accoeding to the thermal analysis, the water cooling is good for the cavity working in the high power. During the beam dynamics simulation, the envelope can be controlled under 20mm in the cavity.

## 8 Acknowledgment

This work is support by National Science and Technology Support Program (Grant No. 2014BAA03B01) and China National Nuclear Corporation (Grant No. FA14000104).